\documentclass[cits]{PoS}

\usepackage{amsmath}
\usepackage{subcaption}
\usepackage[utf8x]{inputenc}

\usepackage{feynmp}
\makeatletter
\def\endfmffile{%
  \fmfcmd{\p@rcent\space the end.^^J%
          end.^^J%
          endinput;}%
  \if@fmfio
    \immediate\closeout\@outfmf
  \fi
  \ifnum\pdfshellescape=\@ne
    \immediate\write18{mpost \thefmffile}%
  \fi}
\makeatother

\title{Lepton flavour universality violation from composite muons}

\ShortTitle{Lepton flavour universality violation from composite muons}

\author{\speaker{Peter Stangl}%
         \thanks{This article is based on \cite{Niehoff:2015bfa} done in collaboration with Christoph Niehoff and David Straub.}
\\
        Excellence Cluster Universe, TUM, Boltzmannstr.~2, 85748~Garching, 
Germany\\
        E-mail: \email{peter.stangl@ph.tum.de}}

\abstract{We describe a possibility to explain the $2.6\sigma$ deviation from lepton flavour universality
observed by the LHCb collaboration in $B^+\to K^+\ell^+\ell^-$ decays in the context of minimal composite Higgs models.
We find that a sizable degree of compositeness of partially composite muons can lead to a good agreement with the
experimental data. Our construction predicts a new physics contribution to $B_s$-$\bar{B}_s$ mixing. Additionally, it accounts for the deficit in the invisible $Z$ width measured at LEP.}

\FullConference{The European Physical Society Conference on High Energy Physics\\
		22--29 July 2015\\
		Vienna, Austria}

\begin{document}

\section{Introduction}

The LHCb collaboration has measured a $2.6\sigma$ deviation from the Standard Model (SM) value
of $R_K$, which is the ratio of the $B^+\to K^+\mu^+\mu^-$ and $B^+\to K^+e^+e^-$ branching ratios \cite{Aaij:2014ora}:
\begin{equation}\label{eq:R_K}
 R_K = \frac{\text{BR}(B^+ \to K^+\mu^+\mu^-)_{[1,6]}}{\text{BR}(B^+ \to 
K^+e^+e^-)_{[1,6]}} = 0.745^{+0.090}_{-0.074} \pm 0.036
 \,.
\end{equation}
In the SM, due to lepton flavour universality (LFU), $R_K$ is to a very good approximation equal
to $1.0$ \cite{Bobeth:2007dw}.
A confirmation of the LHCb measurement would thus be a clear sign of new physics (NP) violating LFU.

There have been several attempts to explain (\ref{eq:R_K}) in terms of NP models. Most models
capable of doing this contain spin-0 or spin-1 leptoquarks or a neutral heavy gauge-boson
mediating the $b\to s\ \ell^+\ell^-$ transition at tree level \cite{Altmannshofer:2014cfa,Buras:2014fpa,Glashow:2014iga,Bhattacharya:2014wla,Crivellin:2015mga,Altmannshofer:2014rta,Crivellin:2015lwa,Hiller:2014yaa,Biswas:2014gga,Sahoo:2015wya,Hiller:2014ula,Sierra:2015fma,Celis:2015ara,Altmannshofer:2015mqa,Falkowski:2015zwa}.
While (\ref{eq:R_K}) can not be reproduced in the Minimal Supersymmetric Standard Model (MSSM) 
\cite{Altmannshofer:2014rta}, it was shown that this is possible in composite Higgs models (CHMs) when introducing composite leptoquarks
\cite{Gripaios:2014tna}.
In \cite{Niehoff:2015bfa}, a mechanism was described how also more simple CHMs without leptoquarks could be used to explain (\ref{eq:R_K}). This approach is presented in more detail in the following sections, where we closely follow \cite{Niehoff:2015bfa}.

\section{Explaining $R_K$ in composite Higgs models}
At the quark-level, the
$B^+\to K^+\ell^+\ell^-$ decay is
based on the $b\to s\ \ell^+\ell^-$ transition.
For its parametrization we consider
the Wilson coefficients $C_{9}^{\ell}$,  ${C^{\,\prime}}_{9}^{\ell}$,  $C_{10}^{\ell}$ and ${C^{\,\prime}}_{10}^{\ell}$ associated with the operators
\begin{align*}
O_{9}^{\ell}&=\left(\bar{s}\,\gamma_{\mu}\,P_{L}\,b\right)\left(\bar{\ell}\,\gamma^{\mu}\,\ell\right),
&\quad
{O'}_{9}^{\ell}&=\left(\bar{s}\,\gamma_{\mu}\,P_{R}\,b\right)\left(\bar{\ell}\,\gamma^{\mu}\,\ell\right),
\\
O_{10}^{\ell}&=\left(\bar{s}\,\gamma_{\mu}\,P_{L}\,b\right)\left(\bar{\ell}\,\gamma^{\mu}\,\gamma_{5}\,\ell\right),
&\quad
{O'}_{10}^{\ell}&=\left(\bar{s}\,\gamma_{\mu}\,P_{R}\,b\right)\left(\bar{\ell}\,\gamma^{\mu}\,\gamma_{5}\,\ell\right),
\end{align*}
which are contained in the weak effective Hamiltonian.
A global analysis on $b \to s$ transitions including several flavour observables \cite{Altmannshofer:2014rta} (see also \cite{Ghosh:2014awa,Hurth:2014vma}) has shown that data prefer either
\begin{itemize}
\item a negative shift in $C_{9}^{\mu}$ only: $\delta C_{9}^{\mu}<0$
\item or a shift in $C_{9}^{\mu}$ and $C_{10}^{\mu}$ with $-\delta C_{10}^{\mu}=\delta C_{9}^{\mu}<0$,
\end{itemize}
where $\delta C_i^{\mu}=C_i^{\mu}-C_i^{\mu\,(\text{SM})}$ denotes the shift in the muonic Wilson coefficients with respect to their values in the SM.

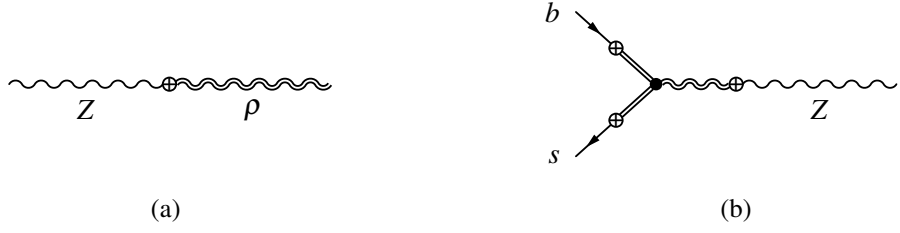
\begin{figure}[t]
\centering
\begin{subfigure}{0.49\textwidth}
\centering
$
 \begin{fmffile}{ZRhoMixing}
\vcenter{\hbox{
\begin{fmfgraph*}(120,70)
\fmfset{arrow_len}{2mm}
\fmfset{thin}{.7pt}
\fmfpen{thin}
\fmfstraight

\fmfleft{x}
\fmfright{y}

\fmf{boson,label=$Z$}{x,s}
\fmf{dbl_wiggly,label=$\rho$}{s,y}

\fmfv{decor.shape=circle,decor.filled=empty,decor.size=5pt,label.dist=0,label=\small{+}}{s}

\end{fmfgraph*}}}
\end{fmffile}
$
\caption{}
\label{fig:mixing}
\end{subfigure}
\begin{subfigure}{0.49\textwidth}
\centering
$
\begin{fmffile}{FCNC}
\vcenter{\hbox{
\begin{fmfgraph*}(120,70)
\fmfset{arrow_len}{2mm}
\fmfset{thin}{.7pt}
\fmfpen{thin}
\fmfstraight

\fmfleft{x1,s,,,,,,,b,x2}
\fmfright{v2}

\fmf{fermion,tension=2}{b,mb}
\fmf{double,tension=2}{mb,v1}
\fmf{fermion,tension=2}{ms,s}
\fmf{double,tension=2}{v1,ms}

\fmf{dbl_wiggly,tension=2}{v1,mZ}
\fmf{boson,tension=1,label=$Z$}{mZ,v2}

\fmfdot{v1}
\fmfv{decor.shape=circle,decor.filled=empty,decor.size=5pt,label.dist=0,label=\small{+}}{ms,mb,mZ}

\fmfv{lab=$b$,l.angle=-180}{b}
\fmfv{lab=$s$,l.angle=-180}{s}

\end{fmfgraph*}
}}
\end{fmffile}
$
\caption{}
\label{fig:fcnc}
\end{subfigure}
\caption{
(a): Mixing of an elementary $Z$ boson with its heavy partner $\rho$. 
(b): Tree level FCNC due to mixing of $b$,$s$ and $Z$ with their heavy partners.}
\end{figure}
To explain a contribution to the above mentioned Wilson coefficients in the context of CHMs, we mainly rely on one
specific property of the bulk of these models, namely the notion of partial compositeness.
Here we present only the most
important concepts that are necessary to describe our construction. For more details on the chosen model we refer to 
\cite{Niehoff:2015bfa}.
In the model we consider, the SM gauge and fermion field content makes up the so called elementary sector.
In addition, there is a composite sector that contains a heavy resonance partner for each of the elementary fields.
These heavy partners
transform under a global symmetry group $G$ that contains the SM gauge group $G_{SM}$ as a subgroup.\footnote{In the following we use $G = SO(5)\times U(1)_X$. This is motivated by models with the Higgs field consisting of the pseudo Nambu-Goldstone bosons
of an $SO(5)\to SO(4)$ global symmetry breaking. The $U(1)_X$ is needed to account for the correct hypercharges. See \cite{Niehoff:2015bfa} for more details.}
Elementary and composite fields are coupled to each other such that they mix (cf. fig. \ref{fig:mixing}).
The amount of mixing
is called the ``degree of compositeness''\footnote{
The mass eigenstates of the theory have to be obtained by a diagonalization of the mass matrices. Due to the mixing, after the diagonalization these eigenstates then contain parts of both elementary and composite fields. 
The lowest mass eigenstates are finally identified with the mass eigenstates of the SM. How much of the composite fields they contain is controlled by the amount of mixing and thus its name ``degree of compositeness''.}.
Due to the mixing, couplings between the elementary fields are modified and new couplings to the heavy sector are introduced.
This leads to e.g. tree level flavour changing neutral currents (FCNCs) that are not present in the SM (cf. fig. \ref{fig:fcnc}).

\begin{figure}[b]
\centering
\begin{subfigure}{0.32\textwidth}
\centering
$
\begin{fmffile}{Zexchange}
\vcenter{\hbox{
\begin{fmfgraph*}(90,70)
\fmfset{arrow_len}{2mm}
\fmfset{thin}{.7pt}
\fmfpen{thin}
\fmfstraight

\fmfleft{x1,s,,,,,,,b,x2}
\fmfright{y1,mum,,,,,,,mup,y2}

\fmf{fermion,tension=2}{b,mb}
\fmf{double,tension=2}{mb,v1}
\fmf{fermion,tension=2}{ms,s}
\fmf{double,tension=2}{v1,ms}

\fmf{dbl_wiggly,tension=2}{v1,mZ}
\fmf{boson,tension=1,label=$Z$}{mZ,v2}

\fmf{fermion}{v2,mup}
\fmf{fermion}{mum,v2}

\fmfdot{v1,v2}
\fmfv{decor.shape=circle,decor.filled=empty,decor.size=5pt,label.dist=0,label=\small{+}}{ms,mb,mZ}

\fmfv{lab=$b$,l.angle=-180}{b}
\fmfv{lab=$s$,l.angle=-180}{s}
\fmfv{lab=$\mu^-$,l.angle=0}{mup}
\fmfv{lab=$\mu^+$,l.angle=0}{mum}

\end{fmfgraph*}
}}
\end{fmffile}
$
\caption{}
\label{fig:Zexchange}
\end{subfigure}
\begin{subfigure}{0.32\textwidth}
\centering
$
\begin{fmffile}{RHOexchange1}
\vcenter{\hbox{
\begin{fmfgraph*}(90,70)
\fmfset{arrow_len}{2mm}
\fmfset{thin}{.7pt}
\fmfpen{thin}
\fmfstraight

\fmfleft{x1,s,,,,,,,b,x2}
\fmfright{y1,mum,,,,,,,mup,y2}

\fmf{fermion,tension=2}{b,mb}
\fmf{double,tension=2}{mb,v1}
\fmf{fermion,tension=2}{ms,s}
\fmf{double,tension=2}{v1,ms}

\fmf{dbl_wiggly,tension=2/3,label=$\rho$}{v1,v2}

\fmf{fermion}{v2,mup}
\fmf{fermion}{mum,v2}

\fmfdot{v1,v2}
\fmfv{decor.shape=circle,decor.filled=empty,decor.size=5pt,label.dist=0,label=\small{+}}{ms,mb}

\fmfv{lab=$b$,l.angle=-180}{b}
\fmfv{lab=$s$,l.angle=-180}{s}
\fmfv{lab=$\mu^-$,l.angle=0}{mup}
\fmfv{lab=$\mu^+$,l.angle=0}{mum}

\end{fmfgraph*}}}
\end{fmffile}
$
\caption{}
\label{fig:rhoexchange1}
\end{subfigure}
\begin{subfigure}{0.32\textwidth}
\centering
$
\begin{fmffile}{RHOexchange2}
\vcenter{\hbox{
\begin{fmfgraph*}(90,70)
\fmfset{arrow_len}{2mm}
\fmfset{thin}{.7pt}
\fmfpen{thin}
\fmfstraight

\fmfleft{x1,s,,,,,,,b,x2}
\fmfright{y1,mum,,,,,,,mup,y2}

\fmf{fermion,tension=2}{b,mb}
\fmf{double,tension=2}{mb,v1}
\fmf{fermion,tension=2}{ms,s}
\fmf{double,tension=2}{v1,ms}

\fmf{dbl_wiggly,tension=2/3,label=$\rho$}{v1,v2}

\fmf{fermion,tension=2}{mmup,mup}
\fmf{double,tension=2}{v2,mmup}
\fmf{fermion,tension=2}{mum,mmum}
\fmf{double,tension=2}{mmum,v2}

\fmfdot{v1,v2}
\fmfv{decor.shape=circle,decor.filled=empty,decor.size=5pt,label.dist=0,label=\small{+}}{ms,mb,mmup,mmum}

\fmfv{lab=$b$,l.angle=-180}{b}
\fmfv{lab=$s$,l.angle=-180}{s}
\fmfv{lab=$\mu^-$,l.angle=0}{mup}
\fmfv{lab=$\mu^+$,l.angle=0}{mum}

\end{fmfgraph*}}}
\end{fmffile}
$
\caption{}
\label{fig:rhoexchange2}
\end{subfigure}
\caption{
(a): $Z$ exchange. 
(b): $\rho$~exchange with $\rho$-muon coupling due to $Z$-$\rho$ mixing.
(c):~$\rho$~exchange with $\rho$-muon coupling due to muons mixing with their heavy partners.
}
\end{figure}
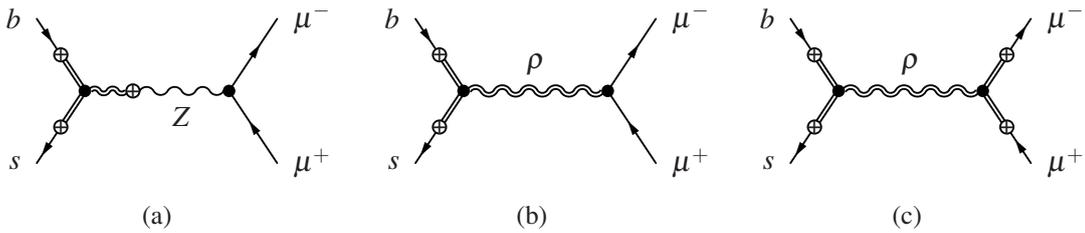

The new and the modified couplings can now be used for trying to get a contribution that fits one of the two cases that are preferred by experimental data ($\delta C_{9}^{\mu}<0$ or $-\delta C_{10}^{\mu}=\delta C_{9}^{\mu}<0$) (cf.~\cite{Straub:2013zca,Altmannshofer:2013foa}).
One possibility is to use the coupling from fig. \ref{fig:fcnc} to get the $Z$ exchange diagram shown in fig. \ref{fig:Zexchange}.
Due to the small vector coupling of the $Z$ to leptons, this diagram gives a contribution with $\delta C_{10}^{\mu}\gg\delta C_{9}^{\mu}$, which is not the desired pattern. In addition, this diagram is lepton flavour universal and thus can not be used to explain (\ref{eq:R_K}).
The second possibility is the exchange of the heavy resonance $\rho$ shown in fig. \ref{fig:rhoexchange1} and \ref{fig:rhoexchange2}.
In the first case, the $\rho$ couples to muons via the mixing of $\rho$ and $Z$. This $\rho$-muon coupling is approximately equal to the coupling of the $Z$ to muons and thus the corresponding diagram suffers from the same problem as the $Z$ exchange.
But the second case of the $\rho$ exchange is different.
Here the muons mix with their heavy partners and they subsequently couple to the $\rho$ via a composite sector coupling. With a sizable degree of compositeness of the muons $s_\mu$, this diagram might give the desired contribution.
For the case of a shift in only $C_9^\mu$, one would need a sizable degree of compositeness for both left- and right-handed muons.
This is problematic because the product of both enters the muon mass and has to be small.
So the first case is thus already ruled out.
The case with $-\delta C_{10}^{\mu}=\delta C_{9}^{\mu}<0$ on the other hand would require only a sizable degree of compositeness of the left-handed muons ${s_{\mu}}_{L}$. This seems possible and calls for further investigation.

\section{Constraints from quark flavour physics and electroweak precision tests}
Assuming that the $\rho$ exchange diagram, where the left-handed muons couple to the $\rho$ via a sizable degree of compositeness (fig. \ref{fig:rhoexchange2}) explains (\ref{eq:R_K}), one has to show that this is not in conflict with other observations.
There are constraints from direct searches for heavy resonances, but in the considered models these resonances are usually heavy enough to avoid them.\footnote{For a comprehensive analysis of composite Higgs models including constraints from direct searches, see e.g. \cite{Niehoff:2015iaa}.}
Getting an effect in $R_K$ requires a flavour changing coupling of $b$ and $s$ quarks to the $\rho$ resonance.
Such a coupling then also leads to a NP contribution to $B_s$-$\bar{B}_s$ mixing which is experimentally constrained \cite{Altmannshofer:2014rta} (see also 
\cite{Altmannshofer:2013foa,Altmannshofer:2009ma,Descotes-Genon:2013wba,Gauld:2013qba,Buras:2013qja}).
Therefore this coupling cannot be too large and its smallness has to be compensated by the sizable degree of compositeness ${s_{\mu}}_{L}$.
Requiring a sizable effect in $R_K$
then leads to the lower bound
\begin{equation}
 {s_{\mu}}_{L}\gtrsim 0.17\cdot\sqrt{f/v},
\end{equation}
where $f$ is the NP scale\footnote{In models were the Higgs is a pseudo Nambu-Goldstone boson, $f$ is the Higgs decay constant. In the model considered here, $f$ is linked to the $\rho$ mass $m_\rho$ by $m_\rho = \tfrac{1}{2}\, g_\rho\,f$, where $g_\rho$ is the coupling between the $\rho$ and the heavy lepton partners.} and $v$ is the Higgs VEV.

A sizable ${s_{\mu}}_{L}$ potentially modifies the couplings of the second generation leptons to the massive electroweak gauge bosons.
A shift in the $Z$ coupling to left-handed muons is very problematic since this coupling is strongly constrained by LEP.
Fortunately, a NP tree level\footnote{The analysis presented here is merely a proof of concept. In a more complete one, an additional loop-correction might be relevant (cf. \cite{Grojean:2013qca}).} contribution to the $Z$-muon coupling can be avoided by a custodial protection through a discrete $P_{LR}$ symmetry \cite{Agashe:2006at,Agashe:2009tu}.
Interestingly, this then already fixes the representations of the global symmetry group $G$ under which the heavy lepton partners transform.
In contrast to the $Z$-muon coupling, the charged current coupling of the $W$ to second generation leptons cannot be protected. This leads to a negative relative shift in the Fermi constant
\begin{equation}
 \frac{\delta G_{F}}{G_{F}}\approx-\frac{v^{2}}{4\,f^{2}}\,{s^{2}_{\mu}}_{L}.
\end{equation}
As it can be seen in fig. \ref{fig:ewpt}, the constraints on $G_F$ are correlated with the constraints on the electroweak $T$ parameter (following \cite{Wells:2014pga}). The maximal shift in $G_F$ that differs by less than $3\sigma$ from the central experimental value gives an upper bound on the degree of compositeness,
\begin{equation}
 {s_{\mu}}_{L}\lesssim 0.08\, \frac{f}{v}.
\end{equation}
Like the charged current coupling, the neutral current coupling to muon neutrinos is also not protected.
This results in a modification of the effective number of light neutrino species $N_\nu$. 
The $2\sigma$ deficit in the invisible $Z$ width measured at LEP \cite{ALEPH:2005ab} can also be expressed by $N_\nu$.
Interestingly, it basically coincides with the value we get from the shift in the coupling between the $Z$ and the muon neutrinos.
So we find that this shift actually improves the agreement with experimental data.

\section{Results}
\begin{figure}[t]
\centering
\begin{subfigure}{0.39\textwidth}
\hspace{-6pt}
\vspace{-73pt}
\includegraphics[width=1.1\textwidth]{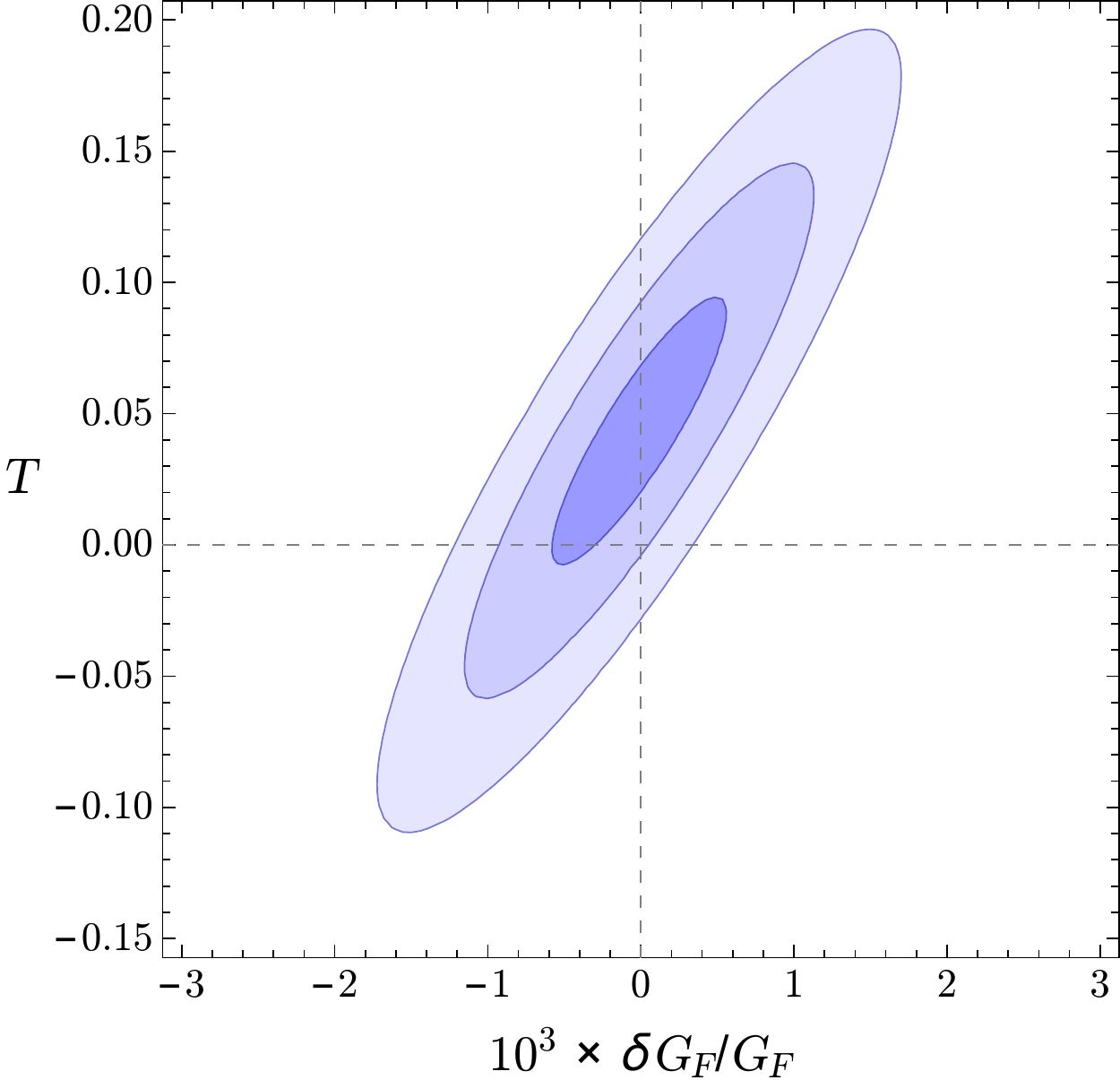}
\vspace{64pt}
\caption{}
\vspace{-64pt}
\label{fig:ewpt}
\end{subfigure}
\begin{subfigure}{0.59\textwidth}
\hspace{10pt}
\includegraphics[width=1\textwidth]{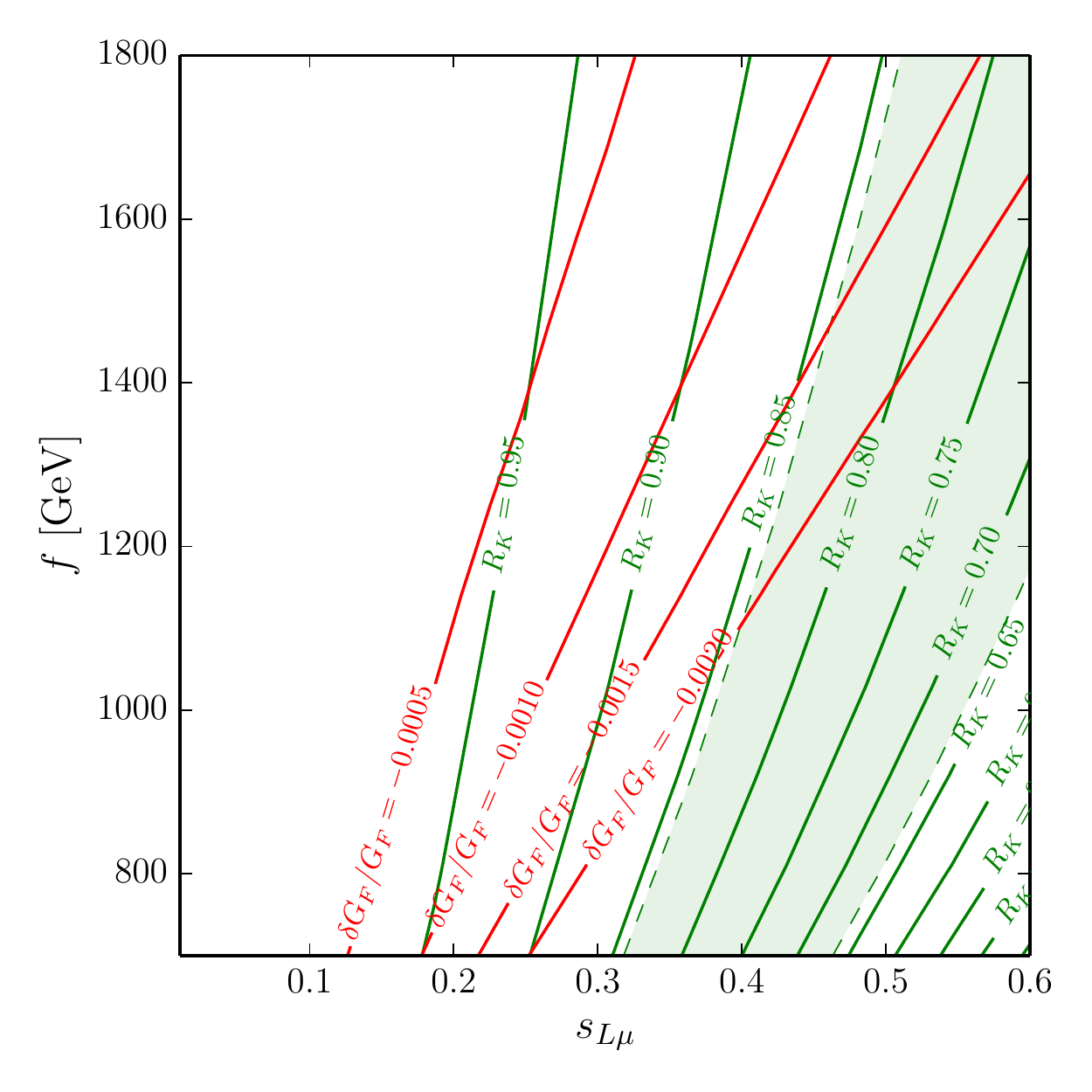}
\vspace{-20pt}
\caption{}
\label{fig:constraints}
\end{subfigure}
\caption{(a): Constraints at the 1, 2 and 3$\sigma$ level on the relative shift in the Fermi constant and a modification of the electroweak $T$ parameter.
(b): Results for $R_K$ in the considered model (green lines). The green shaded area is the 1$\sigma$ region allowed by the LHCb measurement (\protect\ref{eq:R_K}). The red lines correspond to different relative shifts in the Fermi constant.}
\end{figure}
After considering the most important constraints, we find that a sizable ${s_{\mu}}_{L}$
is a possible explanation for the anomaly (\ref{eq:R_K}), while
a small NP contribution to $B_s$-$\bar{B}_s$ mixing is essential to get the desired effect in $R_K$.
Assuming a 10\% correction to the $B_s$ mass difference $\Delta M_s$, one can express $R_K$ by only ${s_{\mu}}_{L}$ and $f$:
\begin{equation}
 1 - R_{K} \approx  0.14 \left[\frac{1.3\,\text{TeV}}{f}\right] \left[\frac{{s_{\mu}}_{L}}{0.4}\right]^2
\end{equation}
Different values for $R_K$ in the considered model are shown in fig. \ref{fig:constraints} together with corrections to the Fermi constant. Using all previous assumptions, we find lower bounds for the degree of compositeness ${s_{\mu}}_{L}$ and the NP scale $f$ such that (\ref{eq:R_K}) can be explained at the $1\sigma$ level:
\begin{equation}
 f \gtrsim 1.3\, \text{TeV},
\quad
{s_{\mu}}_{L} \gtrsim 0.4
\end{equation}
In addition, this can explain the $2\sigma$ deficit in the invisible $Z$ width measured at LEP.

\section{Conclusions}
We have shown that left handed muons with a sizable degree of compositeness can explain the departure from LFU measured by LHCb. There are several implications of this explanation.
\begin{itemize}
 \item It relies on a non-universal coupling of only left-handed muons to the $\rho$ resonance and thus predicts $\delta C_{10}^{\mu}=-\delta C_{9}^{\mu}$.
This can be tested with global fits to measurements of processes involving $b\to s$ transitions.
\item Violation of LFU in other modes is expected.
\item A NP contribution to $B_s$ mixing is predicted that is testable with higher precision of CKM parameters.
\item The heavy $\rho$ resonance should be seen at some point by direct searches, but it may be too heavy to be detected at the LHC.
\end{itemize}

\section*{Acknowledgements}
The author thanks Christoph Niehoff and David Straub for the collaboration on the topic presented here.
This work was supported by the DFG cluster of excellence ``Origin and Structure of the Universe''.

\bibliographystyle{JHEP}
\bibliography{Stangl_EPS15_PoS}

\providecommand{\href}[2]{#2}\begingroup\raggedright\begin{thebibliography}{10}

\bibitem{Niehoff:2015bfa}
C.~Niehoff, P.~Stangl, and D.~M. Straub, \textit{{Violation of lepton flavour
  universality in composite Higgs models}},  {\em Phys. Lett.} \textbf{B747}
  (2015) 182--186,
  [\href{http://arxiv.org/abs/1503.03865}{\texttt{arXiv:1503.03865}}].

\bibitem{Aaij:2014ora}
\textbf{LHCb} Collaboration, R.~Aaij et~al., \textit{{Test of lepton
  universality using $B^{+}\rightarrow K^{+}\ell^{+}\ell^{-}$ decays}},  {\em
  Phys. Rev. Lett.} \textbf{113} (2014) 151601,
  [\href{http://arxiv.org/abs/1406.6482}{\texttt{arXiv:1406.6482}}].

\bibitem{Bobeth:2007dw}
C.~Bobeth, G.~Hiller, and G.~Piranishvili, \textit{{Angular distributions of
  $\bar{B} \to \bar{K} \ell^+\ell^-$ decays}},  {\em JHEP} \textbf{12} (2007)
  040, [\href{http://arxiv.org/abs/0709.4174}{\texttt{arXiv:0709.4174}}].

\bibitem{Altmannshofer:2014cfa}
W.~Altmannshofer, S.~Gori, M.~Pospelov, and I.~Yavin, \textit{{Quark flavor
  transitions in $L_\mu-L_\tau$ models}},  {\em Phys. Rev.} \textbf{D89} (2014)
  095033, [\href{http://arxiv.org/abs/1403.1269}{\texttt{arXiv:1403.1269}}].

\bibitem{Buras:2014fpa}
A.~J. Buras, J.~Girrbach-Noe, C.~Niehoff, and D.~M. Straub, \textit{{$ B\to
  {K}^{\left(\ast \right)}\nu \overline{\nu} $ decays in the Standard Model and
  beyond}},  {\em JHEP} \textbf{02} (2015) 184,
  [\href{http://arxiv.org/abs/1409.4557}{\texttt{arXiv:1409.4557}}].

\bibitem{Glashow:2014iga}
S.~L. Glashow, D.~Guadagnoli, and K.~Lane, \textit{{Lepton Flavor Violation in
  $B$ Decays?}},  {\em Phys. Rev. Lett.} \textbf{114} (2015) 091801,
  [\href{http://arxiv.org/abs/1411.0565}{\texttt{arXiv:1411.0565}}].

\bibitem{Bhattacharya:2014wla}
B.~Bhattacharya, A.~Datta, D.~London, and S.~Shivashankara,
  \textit{{Simultaneous Explanation of the $R_K$ and $R(D^{(*)})$ Puzzles}},
  {\em Phys. Lett.} \textbf{B742} (2015) 370--374,
  [\href{http://arxiv.org/abs/1412.7164}{\texttt{arXiv:1412.7164}}].

\bibitem{Crivellin:2015mga}
A.~Crivellin, G.~D’Ambrosio, and J.~Heeck, \textit{{Explaining
  $h\to\mu^\pm\tau^\mp$, $B\to K^* \mu^+\mu^-$ and $B\to K \mu^+\mu^-/B\to K
  e^+e^-$ in a two-Higgs-doublet model with gauged $L_\mu-L_\tau$}},  {\em
  Phys. Rev. Lett.} \textbf{114} (2015) 151801,
  [\href{http://arxiv.org/abs/1501.00993}{\texttt{arXiv:1501.00993}}].

\bibitem{Altmannshofer:2014rta}
W.~Altmannshofer and D.~M. Straub, \textit{{New physics in $b\rightarrow s$
  transitions after LHC run 1}},  {\em Eur. Phys. J.} \textbf{C75} (2015),
  no.~8 382, [\href{http://arxiv.org/abs/1411.3161}{\texttt{arXiv:1411.3161}}].

\bibitem{Crivellin:2015lwa}
A.~Crivellin, G.~D’Ambrosio, and J.~Heeck, \textit{{Addressing the LHC flavor
  anomalies with horizontal gauge symmetries}},  {\em Phys. Rev.} \textbf{D91}
  (2015), no.~7 075006,
  [\href{http://arxiv.org/abs/1503.03477}{\texttt{arXiv:1503.03477}}].

\bibitem{Hiller:2014yaa}
G.~Hiller and M.~Schmaltz, \textit{{$R_K$ and future $b \to s \ell \ell$
  physics beyond the standard model opportunities}},  {\em Phys. Rev.}
  \textbf{D90} (2014) 054014,
  [\href{http://arxiv.org/abs/1408.1627}{\texttt{arXiv:1408.1627}}].

\bibitem{Biswas:2014gga}
S.~Biswas, D.~Chowdhury, S.~Han, and S.~J. Lee, \textit{{Explaining the lepton
  non-universality at the LHCb and CMS within a unified framework}},  {\em
  JHEP} \textbf{02} (2015) 142,
  [\href{http://arxiv.org/abs/1409.0882}{\texttt{arXiv:1409.0882}}].

\bibitem{Sahoo:2015wya}
S.~Sahoo and R.~Mohanta, \textit{{Scalar leptoquarks and the rare $B$ meson
  decays}},  {\em Phys. Rev.} \textbf{D91} (2015), no.~9 094019,
  [\href{http://arxiv.org/abs/1501.05193}{\texttt{arXiv:1501.05193}}].

\bibitem{Hiller:2014ula}
G.~Hiller and M.~Schmaltz, \textit{{Diagnosing lepton-nonuniversality in $b \to
  s \ell \ell$}},  {\em JHEP} \textbf{02} (2015) 055,
  [\href{http://arxiv.org/abs/1411.4773}{\texttt{arXiv:1411.4773}}].

\bibitem{Sierra:2015fma}
D.~Aristizabal~Sierra, F.~Staub, and A.~Vicente, \textit{{Shedding light on the
  $b\to s$ anomalies with a dark sector}},  {\em Phys. Rev.} \textbf{D92}
  (2015), no.~1 015001,
  [\href{http://arxiv.org/abs/1503.06077}{\texttt{arXiv:1503.06077}}].

\bibitem{Celis:2015ara}
A.~Celis, J.~Fuentes-Martin, M.~Jung, and H.~Serodio, \textit{{Family
  nonuniversal Z′ models with protected flavor-changing interactions}},  {\em
  Phys. Rev.} \textbf{D92} (2015), no.~1 015007,
  [\href{http://arxiv.org/abs/1505.03079}{\texttt{arXiv:1505.03079}}].

\bibitem{Altmannshofer:2015mqa}
W.~Altmannshofer and I.~Yavin, \textit{{Predictions for Lepton Flavor
  Universality Violation in Rare B Decays in Models with Gauged $L_\mu -
  L_\tau$}},
  \href{http://arxiv.org/abs/1508.07009}{\texttt{arXiv:1508.07009}}.

\bibitem{Falkowski:2015zwa}
A.~Falkowski, M.~Nardecchia, and R.~Ziegler, \textit{{Lepton Flavor
  Non-Universality in B-meson Decays from a U(2) Flavor Model}},
  \href{http://arxiv.org/abs/1509.01249}{\texttt{arXiv:1509.01249}}.

\bibitem{Gripaios:2014tna}
B.~Gripaios, M.~Nardecchia, and S.~A. Renner, \textit{{Composite leptoquarks
  and anomalies in $B$-meson decays}},  {\em JHEP} \textbf{05} (2015) 006,
  [\href{http://arxiv.org/abs/1412.1791}{\texttt{arXiv:1412.1791}}].

\bibitem{Ghosh:2014awa}
D.~Ghosh, M.~Nardecchia, and S.~A. Renner, \textit{{Hint of Lepton Flavour
  Non-Universality in $B$ Meson Decays}},  {\em JHEP} \textbf{12} (2014) 131,
  [\href{http://arxiv.org/abs/1408.4097}{\texttt{arXiv:1408.4097}}].

\bibitem{Hurth:2014vma}
T.~Hurth, F.~Mahmoudi, and S.~Neshatpour, \textit{{Global fits to $b \to
  s\ell\ell$ data and signs for lepton non-universality}},  {\em JHEP}
  \textbf{12} (2014) 053,
  [\href{http://arxiv.org/abs/1410.4545}{\texttt{arXiv:1410.4545}}].

\bibitem{Straub:2013zca}
D.~M. Straub, \textit{{Anatomy of flavour-changing $Z$ couplings in models with
  partial compositeness}},  {\em JHEP} \textbf{08} (2013) 108,
  [\href{http://arxiv.org/abs/1302.4651}{\texttt{arXiv:1302.4651}}].

\bibitem{Altmannshofer:2013foa}
W.~Altmannshofer and D.~M. Straub, \textit{{New physics in $B \to
  K^*\mu\mu$?}},  {\em Eur. Phys. J.} \textbf{C73} (2013) 2646,
  [\href{http://arxiv.org/abs/1308.1501}{\texttt{arXiv:1308.1501}}].

\bibitem{Niehoff:2015iaa}
C.~Niehoff, P.~Stangl, and D.~M. Straub, \textit{{Direct and indirect signals
  of natural composite Higgs models}},
  \href{http://arxiv.org/abs/1508.00569}{\texttt{arXiv:1508.00569}}.

\bibitem{Altmannshofer:2009ma}
W.~Altmannshofer, A.~J. Buras, D.~M. Straub, and M.~Wick, \textit{{New
  strategies for New Physics search in $B \to K^{*} \nu \bar{\nu}$, $B \to K
  \nu \bar{\nu}$ and $B \to X_{s} \nu \bar{\nu}$ decays}},  {\em JHEP}
  \textbf{04} (2009) 022,
  [\href{http://arxiv.org/abs/0902.0160}{\texttt{arXiv:0902.0160}}].

\bibitem{Descotes-Genon:2013wba}
S.~Descotes-Genon, J.~Matias, and J.~Virto, \textit{{Understanding the $B\to
  K^*\mu^+\mu^-$ Anomaly}},  {\em Phys. Rev.} \textbf{D88} (2013) 074002,
  [\href{http://arxiv.org/abs/1307.5683}{\texttt{arXiv:1307.5683}}].

\bibitem{Gauld:2013qba}
R.~Gauld, F.~Goertz, and U.~Haisch, \textit{{On minimal $Z'$ explanations of
  the $B\to K^*\mu^+\mu^-$ anomaly}},  {\em Phys. Rev.} \textbf{D89} (2014)
  015005, [\href{http://arxiv.org/abs/1308.1959}{\texttt{arXiv:1308.1959}}].

\bibitem{Buras:2013qja}
A.~J. Buras and J.~Girrbach, \textit{{Left-handed $Z'$ and $Z$ FCNC quark
  couplings facing new $b \to s \mu^+ \mu^-$ data}},  {\em JHEP} \textbf{12}
  (2013) 009,
  [\href{http://arxiv.org/abs/1309.2466}{\texttt{arXiv:1309.2466}}].

\bibitem{Grojean:2013qca}
C.~Grojean, O.~Matsedonskyi, and G.~Panico, \textit{{Light top partners and
  precision physics}},  {\em JHEP} \textbf{10} (2013) 160,
  [\href{http://arxiv.org/abs/1306.4655}{\texttt{arXiv:1306.4655}}].

\bibitem{Agashe:2006at}
K.~Agashe, R.~Contino, L.~Da~Rold, and A.~Pomarol, \textit{{A Custodial
  symmetry for $Zb \bar b$}},  {\em Phys. Lett.} \textbf{B641} (2006) 62--66,
  [\href{http://arxiv.org/abs/hep-ph/0605341}{\texttt{hep-ph/0605341}}].

\bibitem{Agashe:2009tu}
K.~Agashe, \textit{{Relaxing Constraints from Lepton Flavor Violation in 5D
  Flavorful Theories}},  {\em Phys. Rev.} \textbf{D80} (2009) 115020,
  [\href{http://arxiv.org/abs/0902.2400}{\texttt{arXiv:0902.2400}}].

\bibitem{Wells:2014pga}
J.~D. Wells and Z.~Zhang, \textit{{Precision Electroweak Analysis after the
  Higgs Boson Discovery}},  {\em Phys. Rev.} \textbf{D90} (2014), no.~3 033006,
  [\href{http://arxiv.org/abs/1406.6070}{\texttt{arXiv:1406.6070}}].

\bibitem{ALEPH:2005ab}
\textbf{SLD Electroweak Group, DELPHI, ALEPH, SLD, SLD Heavy Flavour Group,
  OPAL, LEP Electroweak Working Group, L3} Collaboration, S.~Schael et~al.,
  \textit{{Precision electroweak measurements on the $Z$ resonance}},  {\em
  Phys. Rept.} \textbf{427} (2006) 257--454,
  [\href{http://arxiv.org/abs/hep-ex/0509008}{\texttt{hep-ex/0509008}}].

\end{thebibliography}\endgroup

\end{document}